\documentclass[11pt]{article}
\usepackage{epsfig}
\usepackage{graphicx,graphics}
\usepackage{amssymb,amsmath}
\usepackage{verbatim,enumerate}
\usepackage{rotating}
\usepackage{verbatim,multicol,color}
\usepackage{lmodern}
\usepackage{array}
\usepackage{setspace}
\usepackage{multirow}
\usepackage{tabularx}
\usepackage{lscape, pdflscape}
\usepackage{url}
\usepackage{dcolumn}
\newcolumntype{d}{D{.}{.}{2.5}}
\newcolumntype{s}{D{.}{.}{1.2}}
\RequirePackage{lineno}

\def\log{\hbox{log}}

\def\boxit#1{\vbox{\hrule\hbox{\vrule\kern6pt
          \vbox{\kern6pt#1\kern6pt}\kern6pt\vrule}\hrule}}
\def\refhg{\hangindent=20pt\hangafter=1}
\def\refmark{\par\vskip 2mm\noindent\refhg}

\def\refhg{\hangindent=20pt\hangafter=1}
\def\refmark{\par\vskip 2mm\noindent\refhg}

\def\bse{\begin{eqnarray*}}
\def\ese{\end{eqnarray*}}
\def\be{\begin{eqnarray}}
\def\ee{\end{eqnarray}}
\def\bq{\begin{equation}}
\def\eq{\end{equation}}
\def\bse{\begin{eqnarray*}}
\def\ese{\end{eqnarray*}}

\renewcommand{\baselinestretch}{1.5} 

\newtheorem{Th}{Theorem}

\renewcommand{\d}{\stackrel{d}{\rightarrow}}

\def\part{\partial}

\renewcommand{\baselinestretch}{1}

\renewcommand{\d}{\,\mathrm{d}}

\begin{document}

{\center\LARGE  On the spatial correlation between areas of high coseismic slip and aftershock\\ clusters of the Maule earthquake $M_w$=8.8\\}
\vskip15mm

{\center {\bf Javier E. Contreras-Reyes}\\
\vskip2mm
Departamento de Estad\'istica\\

Universidad de Valpara\'iso\\

Valpara\'iso, Chile\\

Email: jecontrr@mat.puc.cl\\}
\vskip8mm
{\center {\bf Adelchi Azzalini}\\
\vskip2mm
Dipartimento di Scienze Statistiche\\

Universit\`a di Padova\\

Padova, Italia\\

Email: azzalini@stat.unipd.it\\}
\vskip15mm

\begin{abstract}
  We study the spatial distribution of clusters associated to the aftershocks of the
  megathrust Maule earthquake $M_W$ 8.8 of 27 February 2010. We used a recent clustering method
   which hinges on a nonparametric estimation of the
  underlying probability density function to detect subsets of points forming
  clusters associated with high density areas.  In addition, we
  estimate the probability density function using a nonparametric kernel
  method for each of these clusters. This allows us to
  identify a set of regions where there is an association between
  frequency of events and coseismic slip. Our results
  suggest that high coseismic slip spatially correlates with high aftershock frequency.\\


  {\it Short title:} Nonparametric assessment of Maule earthquake.
\end{abstract}

\vskip10mm

\section{Introduction}\label{sec.intro}

The most recent large Chilean
earthquake occurred on February 27th, 2010 ($M_w$ = 8.8) and
propagated northward and southward achieving a rupture
length of about 450 km (35--$37^{\circ}$S; e.g., Lay et al., 2010). The earthquake
filled a seismic gap (Ruegg et al., 2009) that has experienced little
seismic activity since 1835, when it broke with an estimated
magnitude of $M_w \sim$8.5 (Darwin, 1851).

Moreno et al. (2010) showed that the two regions of high coseismic slip of
the Maule earthquake appeared to be highly
locked before the earthquake. 
Subsequent geodetic studies have established that the
main coseismic slip patch ($>$15 m) is located in the northern part of
the rupture area, with a secondary concentration of slip to the south
(5-12 m) (e.\,g.\ Lay et al., 2010; Delouis et al., 2010; Lorito
et al. 2011; Vigny et al., 2011; Moreno et al., 2012). Lorito et al.
(2011) concluded that increased stress on the unbroken southern
patch may have increased the probability of another great earthquake there in the near
future, but his model has poor resolution on this area. In addition,
Moreno et al. (2012) suggests that coseismic slip heterogeneity
at the scale of single asperities appear to indicate seismic potential for
future great earthquakes. These studies are not limited to \emph{ geodetic data},
seismic and tsunami data have been used as well in those studies.

The aim of the present work is to examine the spatial correlation between areas
of high coseismic slip and the aftershock frequency (seismic clusters).
To achieve this, we identify the
distribution of clusters in the rupture area of the 2010 earthquake. The hypocenter
data are taken from the SSN (Servicio Sismologico Nacional de Chile), and we use
the coseismic model of Moreno et al. (2012), which includes all available
geodetic data for the Maule earthquake. We adopted the NPC formulation of Azzalini \& Torelli (2007), which
has the advantage of not requiring the input of some subjective choices
such as the number of existing clusters. We apply this methodology to
the SSN aftershock catalogue data of the Maule earthquake and focus
on the data of the area between Valpara\'iso and Tir\'ua 
[33-38$\mbox{.}5^{\circ}$S], aiming the identification of
seismic clusters and its spatial relationship with regions of coseismic slip. Finally, we
discuss the possible improvements of the adopted methodology in terms of available geophysical data.

\section{Methodology}

Applications of clustering techniques
range over an enormous set of disciplines, both in the natural sciences (see e.g, Contreras-Reyes \& Arellano-Valle, 2012)
and in the social sciences (see e.g, Azzalini \& Torelli, 2007; Menardi, 2010). A standard account is the book of Kaufman
\& Rousseeuw (1990),  which is focused mainly on the more classical
methods, based of the notion of {\emph distance} between objects.
An alternative, more recent approach is the model-based clustering
formulation which regards  the observed data as generated by a
probability distribution a mixture  of multivariate random variables
having distribution belonging to  some parametric family.
More specifically, for any point $x$ in the multivariate set
of possible observations, we associate a probability density
function $f(x)$ and assume that a decomposition of type
$$f(x)= \sum_{j=1}^{J} p_j f_j(x)$$
exists, where the $p_j$'s are mixing
probabilities and $f_j(x)$ denotes the $j$-th component of the
mixture, which also corresponds to the $j$-th cluster of points,
for $j=1,\dots,J$.
In the implementation of this approach, the most common option is
to adopt the multivariate Gaussian assumption for each of the
components $f_j(x)$,
and estimate their parameters using an EM-type algorithm.
An especially useful account to this approach
is provided by Dasgupta \& Raftery (1998). 
By applying the model-based clustering
approach to earthquake data in the coastal area of central
California, Dasgupta \& Raftery (1998) have obtained six clusters, some of which
are clearly linked to active faults, such as San Andreas, Calveras and Hayward faults;
however one or two of their clusters do not correspond to some already
identified area.

In a further approach to the clustering problem, the notion of an
underlying density function $f(x)$ is retained, but the assumption
that $f(x)$ is a mixture of components is removed, and so also the
connected parametric assumption of the components. Therefore, this
approach is based on the construction of a nonparametric estimate
$\hat{f}(x)$ of $f(x)$, and the association of a cluster to each
of the observed modes of the density. Several alternative formulations
of this approach exists (see e.g., Botev et al., 2010; Zhu, 2013),
but we concentrate on the kernel-type estimate which
is the most popular and conceptually very intuitive (Bowman
\& Azzalini, 1997). In the univariate case, the kernel estimate
is defined by
\begin{equation}\label{KE}
       \hat{f}(x)=\frac{1}{n}\sum_{i=1}^{n}\varphi(x-x_i;h),
\end{equation}
where $\varphi(z;h)$ denotes the normal density function with
mean $0$ and standard deviation $h$ evaluated at point $z$.
The  normal density is adopted for
simplicity and it could be replaced by another density symmetric
about $0$  without much effect on the outcome. A much more important
role is played  by $h$ which in this context is called `smoothing
parameter'. In the $d$-dimensional case, $\varphi(z;h)$ is replaced by
a multivariate density with a $0$ mean vector; the simplest and most
popular choice is the product of $d$ such terms, with a different
smoothing parameter for each of the $d$ components.
Clusters are then formed by the data points associated to
the modes of  $\hat{f}(x)$, and the clusters are separated by
regions of low density of points.
This logic procedure is referred as a nonparametric clustering (NPC).

\subsection{Nonparametric clustering}

We describe briefly the NPC method proposed by
Azzalini \& Torelli (2007). This works by assuming that the available set of
$d$-dimensional observations $S=\{x_1,\dots, x_n\}$ represent a set of points
drawn from a continuous multivariate random variable having an unknown
probability density function $f(x)$, $x\in\mathbb{R}^d$.  In the case we are
concerned with, $x=(\mathrm{latitude}, \mathrm{longitude})$ denotes a
geographical position, hence $d=2$; the data points $x_1,\dots, x_n$ represent
the positions of the observed seismic events.

For any given constant $\alpha$ such that $0\leq \alpha \leq \max_{x}f(x)$,
consider the high density region defined by
\begin{equation} \label{e:R.alpha}
 R_\alpha=\{x:\mbox{ }x\in\mathbb{R}^2,\mbox{ }f(x)\ge\alpha\}
\end{equation}
which has an associated probability $p_\alpha=\int_{R_\alpha} f(x)\d{x}$.
The region $R_\alpha$ is, in general, formed by a number $m$
of connected sets, where $m\in\{1, 2, \dots\}$.

Now we let $\alpha$ moves along its range. This causes both $m$
and $p=p(\alpha)$ to move accordingly, and we can regard $m$ as a function
of $p$ since $p$ is monotonic with respect to $\alpha$, write $m(p)$.
Note that $m(p)$ is instead a not monotonic function.
With the additional conventional settings $m(0)=m(1)=0$,
it can be shown that the total number of increments of the step function
$m(p)$ is equal  to the number $M$ of modes of $f(x)$, hence to the
number of clusters,  in the sense defined earlier.
As $\alpha$ varies along its range, and so does $p(\alpha)$, the
corresponding connected components of $R(\alpha)$ form a hierarchical
tree structure.

Translating the above idea into a working methodology requires some
additional specifications and algorithmic work. We sketch here the
main step of the procedure. Full details of the method are given by
Azzalini \& Torelli (1997) and its implementation in the language R is
provided by Azzalini \emph{et al.} (2011). This procedure will
later referred to as the `pdfCluster method'.

\begin{enumerate}
\item A nonparametric estimate $\hat{f}(x)$ of the density is obtained
from the observed sample $S$. Any sort of nonparametric
estimate can be employed but in practice both by Azzalini \& Torelli (1997) and pdfCluster
adopt a kernel-type estimate. This involves to choose a set of kernel
bandwidths, one for each component of $x$. When the target is the
estimation of the density itself, the outcome depends
critically on the choice of  the bandwidth. However in the present
context, where $\hat{f}(x)$ is only an intermediate step toward the
final target of clustering, the bandwidth is much less influential
and its choice can be varied over a quite wide range without affecting
the end result.

\item For any $\alpha$ in the range $0\leq \alpha \leq \max_i\hat{f}(x_i)$,
it considers the sample analogue of (\ref{e:R.alpha}) given by
\begin{equation} \label{e:S.alpha}
    S_\alpha= \{x_i:~x_i\in S, ~\hat{f}(x_i)\ge\alpha\}\,.
\end{equation}
The notion of connected subsets of $S_\alpha$ is introduced via the
notion of Delaunay triangulation which is built by making use of
computational geometry tools. This leads to establish a certain set
of line segments joining some of the data points in such a way that,
for any  two data  points, there always exists a path joining them
by a sequence of these segments. If  $x_r$ and $x_s$ are both points
in  $S_\alpha$ and the path joining them does not include any point
outside $S_\alpha$,  then we say that $x_r$ and $x_s$ are connected,
at the given choice of $\alpha$.

\item The above step is replicated for a grid of values spanning
the admissible range of $\alpha$, from $0$ to $\max_i\hat{f}(x_i)$.
This generates a mode function $\hat{m}(p)$ and a tree structure
of the  modes.
Notice however that the Delaunay triangulation needs to be
determined only once for all $\alpha$'s.

\item At the end of the earlier step we have obtained a tree structure
of the modes of $\hat{f}(x)$. Moreover, for each of the $M$ modes, we
have allocated some elements of the sample $S$ to the given mode.
These $M$ subsets of data points form the `cluster cores' to which the
remaining points, not belonging to any cluster core, must be aggregated.
For each unallocated point $x_0$, we must select one of distributions
$\hat{f}_1(x), \dots, \hat{f}_M(x)$
which represent the estimated densities of the cluster cores.
This allocation is most naturally based on the likelihood ratio,
that is we allocate $x_0$ to the $j$-th cluster core such that the ratio
\[
   \frac{\hat{f}_j(x_0)}{\max_{k\neq j}\hat{f}_k(x_0)}
\]
is highest. In practice, there are some variant options, depending
on whether the $\hat{f}_j(x)$ densities of the cluster cores are
updated at each new allocation or not, or according to some
intermediate strategy.
\end{enumerate}
The final outcome of the clustering process is represented by the
a partition of the sample $S$ into a set of clusters,
say $C_1, \dots, C_M$.

\subsection{Density-based silhouette diagnostic}

In cluster analysis, the term `silhouette' refers to a diagnostic
tool for the validation of the outcome of the clustering process
(Rousseeuw, 1987). This technique provides a   graphical
representation of how well each object spatially lies within a cluster
to which it has been allocated; hence it provides an indication
of how appropriately the data has been clustered.
The idea arises from the comparison of the small distance of each
observation to the cluster where it has been allocated and a
measure of separation from the closest alternative cluster.

Since the original silhouette is based on a distance measure between
the observations, it is not adequate for NPC methods where distances
do not play an explicit role. An adaption of the silhouette idea to density-based
clustering methods has been proposed by Menardi (2010).
The method, called {\it density-based silhouette} (DBS) diagnostic,
is based on the posterior probabilities
\[
  p_j(x_i)=\frac{\hat\pi_j \hat{f}_j(x_i)}%
             {\sum_{j=1}^{M}\hat\pi_j \hat{f}_j(x_i)}, \qquad
              j=1,\dots,M,
\]
where $\hat\pi_j$  plays the role of prior probability of cluster $C_j$;
in practice it  is taken to be the proportion of points allocated to $C_j$.
The DBS index of observation $x_i$ is
$$\mathrm{dbs}(x_i)=
  \frac{\log\left(\frac{p_{j_0}(x_i)}{p_{j_1}(x_i)}\right)}%
 {\max_{k=1,...,n}\left|\log\left(\frac{p_{j_0}(x_k)}{p_{j_1}(x_k)}
  \right)\right|}
$$
where $j_0$ denotes the cluster to which $x_i$ has been allocated,
and $j_1$ refers to the alternative cluster index for which $p_j$
is maximum,  $j\neq j_0$.

In our case, after partitioning the SSN data with
the pdfCluster method, we applied the DBS diagnostic to
assess the quality of the outcome.

\subsection{Temporal analysis}

We briefly describe four indexes to illustrate the consistency of the NPC
method across the time. These indexes
perform accuracy of the observed in
predicting the corrected category, relative to that of random chance. In
Table~\ref{conting}, $n(F_i,O_j)$ denotes the number of forecasts in category
$i$ that had observations in category $j$, $N(F_i)$ denotes the total number
of forecasts in category $i$, $N(O_j)$ denotes the total number of
observations in category $j$, $N$ is the total number of forecasts and
$i,j=1,...,5$ are the indexes of the five identified clusters.
\begin{eqnarray*}
 NSS &=& \frac{1}{N}\sum_{i=1}^{5}N(F_i,O_i) ,\\
 HSS &=& \frac{NSS-\frac{1}{N^2}\sum_{i=1}^{5}
         N(F_i)N(O_i)}{1-\frac{1}{N^2}\sum_{i=1}^{5}N(F_i)N(O_i)}, \\
 HK &=&\frac{N^2-\sum_{i=1}^{5} N(F_i)N(O_i)}{N^2-\sum_{i=1}^{5}N(O_i)^2}HSS\\
\end{eqnarray*}
where $NSS$ corresponds to normal skill score with range $[0,1]$,
such that 0 indicates no skill, and 1 indicates perfect score.
The index $HSS$ correspond to Heidke skill score (Brier \& Allen, 1952) with range
$(-\infty,1]$, 0 indicates no skill, and 1 indicates perfect score.

\begin{table}
\begin{footnotesize}
\begin{center}
\caption{Multi-category Contingence table with five clusters.}\label{conting}
\vspace{0.5cm}
\begin{tabular}{cccccccc}
  \hline
  \multicolumn{8}{c}{Observed Category}\\
  \hline
  &  & 1 & 2 & 3 & 4 & 5 & Total \\
  \hline
  & 1 & $n(F_1,O_1)$ & $n(F_1,O_2)$ & $n(F_1,O_3)$ & $n(F_1,O_4)$ & $n(F_1,O_5)$ & $N(F_1)$ \\
  & 2 & $n(F_2,O_1)$ & $n(F_2,O_2)$ & $n(F_2,O_3)$ & $n(F_2,O_4)$ & $n(F_2,O_5)$ & $N(F_2)$ \\
  Forecast & 3 & $n(F_3,O_1)$ & $n(F_3,O_2)$ & $n(F_3,O_3)$ & $n(F_3,O_4)$ & $n(F_3,O_5)$ & $N(F_3)$ \\
  Category & 4 & $n(F_4,O_1)$ & $n(F_4,O_2)$ & $n(F_4,O_3)$ & $n(F_4,O_4)$ & $n(F_4,O_5)$ & $N(F_4)$ \\
  & 5 & $n(F_5,O_1)$ & $n(F_5,O_2)$ & $n(F_5,O_3)$ & $n(F_5,O_4)$ & $n(F_5,O_5)$ & $N(F_5)$ \\
  \hline
  Total & & $N(O_1)$ & $N(O_2)$ & $N(O_3)$ & $N(O_4)$ & $N(O_5)$ & $N$ \\
  \hline
\end{tabular}
\end{center}
\end{footnotesize}
\end{table}

The index $HK$ correspond to Hanssen \& Kuipers
discriminant (Hanssen \& Kuipers, 1965) with
range $[-1,1]$, such that 0 indicates no skill, and 1 indicates
perfect score. Similar to the Heidke skill score, except that in the
denominator the fraction of correct forecasts due to random chance
is for an unbiased forecast. Hubert \& Arabie (1985) noticed that
the Rand index is not corrected for chances that are equal to zero
and for random partitions having the same number of objects in each
class. They introduced the corrected Rand index, whose expectation
is equal to zero. The adjusted Rand index is based on three
values: the number $r$ of common joined pairs in O and F, the
expected value $E(r)$ and the maximum value $\max(r)$ of this
index, among the partitions of the same type as O and F. It leads
to the formula
$$HA(O,F)=\frac{r-E(r)}{\max(r)-E(r)}$$
where
\begin{eqnarray*}
r =\sum_{i=1}^{5}\sum_{j=1}^{5}\frac{n_{i,j}(n_{i,j}-1)}{2}, &&
E(r) = \frac{1}{2}\frac{|N(O)|\times|N(F)|}{n(n-1)}\,,\\
|N(O)| = \sum_{i=1}^{5}\frac{n_{i}(n_{i}-1)}{2}, &&
|N(F)| = \sum_{j=1}^{5}\frac{n_{j}(n_{j}-1)}{2}
\end{eqnarray*}
and $\max(r)=\frac{1}{2}(|N(O)|+|N(F)|)$. This maximum value is
questionable since the number of common joined pairs is necessarily bounded by
$\inf{\{|N(O)|,|N(F)|\}}$, but $\max(r)$ insures that the maximum value of
$HA$ is 1 when the two partitions are identical.

\section{Numerical Analysis}

Our numerical work is based on data extracted from the SSN catalogue,
available at \url{http://ssn.dgf.uchile.cl/}. Specifically, we have
considered 6,714 aftershocks in a map [32--$40^{\circ}$S]$\times$[69--$75\mbox{.}5^{\circ}$E],
for a period between 27 February 2010 and 13 July 2011 (see Fig.~\ref{clusters}) and for
local magnitudes $M_l\geq 2\mbox{.}0$. All of these observations have been previous
processed by the SSN with SEISAN~8.3 software
considering the information provided by 22 stations 
located in a map with coordinates $[-33\mbox{.}32, -39\mbox{.}80]$ latitude and
$[-70\mbox{.}29, -73\mbox{.}24]$ longitude.

For the numerical processing, we used the R computing
environment software (R Development Core Team, 2012). Most of the work was done using
the R package {\tt pdfCluster} by Azzalini et al. (2011); this package
comprises the functions {\tt pdfCluster} for the NPC method, {\tt dbs} for DBS
diagnostics and {\tt kepdf} for kernel density estimation.  In addition, R
provides {\tt kruskal.test} function for a nonparametric-type analysis of
variance and {\tt wilcox.test} function for individual test of means.

\subsection{Clustering process}

\begin{figure}[t!]
     \centering
     \includegraphics[width=10cm,height=14cm]{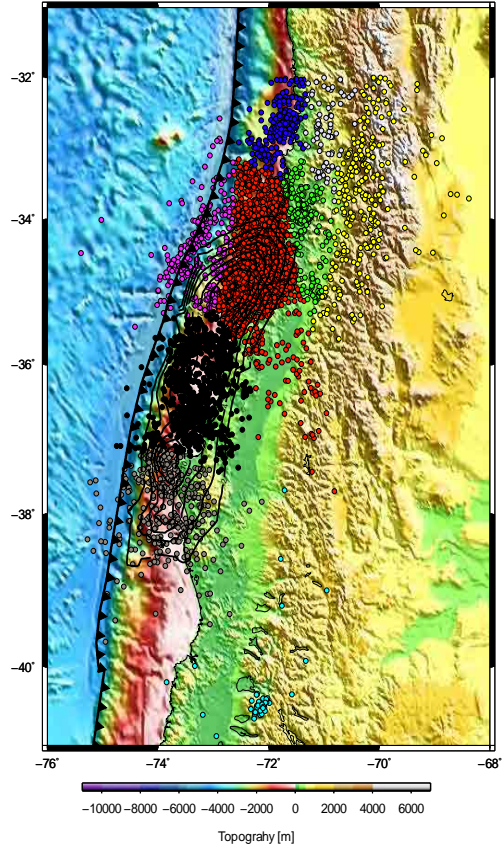}
     \caption{Map of the Chile region analyzed for slip and post-seismicity
       correlation with clustering events. The black-triangle line correspond to the trench.}\label{map}
\end{figure}

Figure~\ref{map} displays the geographical map
of the area of interest with the points denoting the locations of the
events. These points have been clustered using the pdfCluster method
described in Section~2, leading to the different coloring. The bottom
portion of the figure shows cluster tree and the silhouette diagnostic.
These indicate a lack of clear separation among clusters, which is
not surprising, given the  close geographical proximity of the
clouds of points visible in the top panel of the figure.

In the second stage of our numerical work, we have introduced two variants.
One was to consider only the areas were slip did take place in a
non-negligible form, since our aim was exactly to examine the implications of
the slip model. In addition, some numerical exploration has indicated that
log-transformation of the quantities, slip and density function, lead to a more
meaningful outcome. Since for a number of points the slip value is $0$, we
adopted the modification commonly used in this case of adding a small positive
quantity, that is working $\log(k+\mathrm{slip})$, where $k$ is some small
quantity.  Since slip is measured in meters, then we adopted $k=0.01$ which
represents a perturbation of only 1\,cm of the original data.  In the
following, the term log-slip will be used for referring to
$\log(0.01+\mathrm{slip})$.

\begin{figure}[t!]
\includegraphics[width=12cm,height=6.5cm]{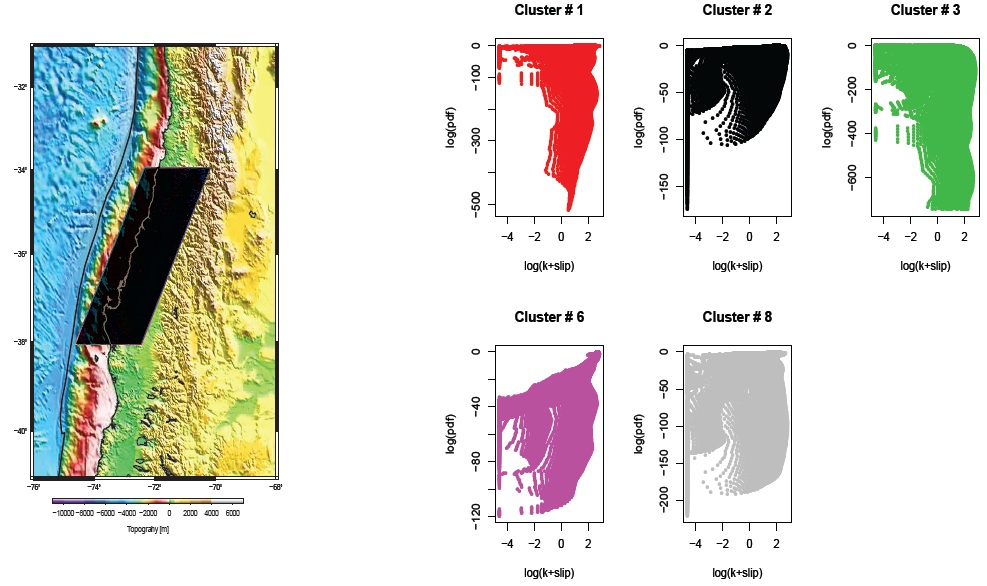}\\
\includegraphics[width=12cm,height=6.5cm]{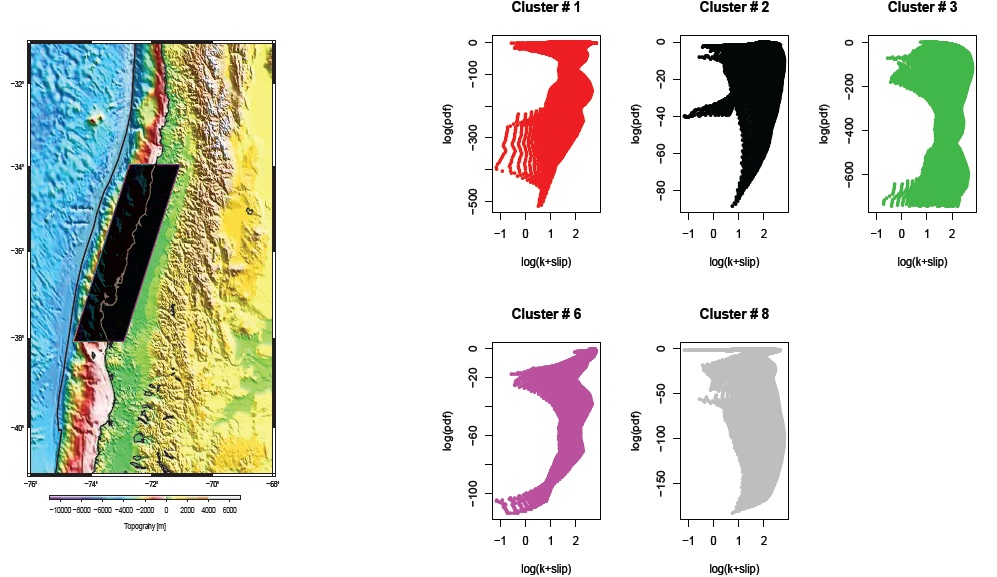}\\
\includegraphics[width=12cm,height=6.5cm]{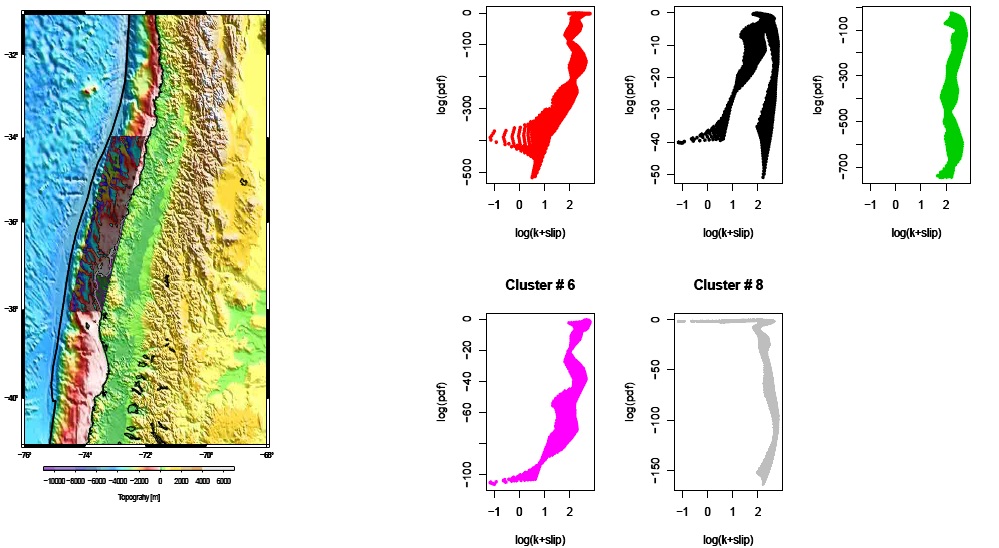}
\caption{Left: Three grids considered for the NPC method (blacked shadows). Right:
     Relationship between log-slip and post-seismicity log-frequency for each
     grid.  The points with $\mathrm{slip}=0$ produce verticals strips of
     points at abscissa $\log(0.01)\approx -4.6$ in some of the top plots.}
  \label{grids}
\end{figure}

Furthermore, association between slip and events density can be examined
in two different ways. One is to chose a regular grid of points in the
region of interest, and evaluate these variables or their log-transforms
over this grid. The other option is to evaluate these variables at the
observed points of the seismic events.

\begin{figure}[htb]
    \centerline{
    \includegraphics[width=0.49\hsize]{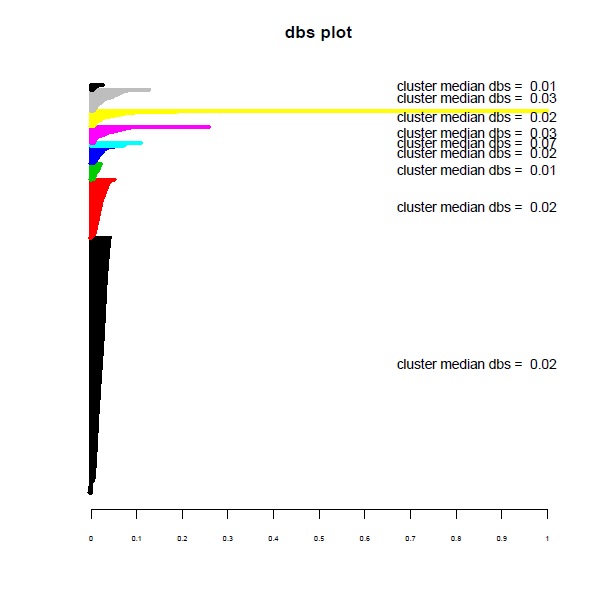}\hfill
    \includegraphics[width=0.49\hsize]{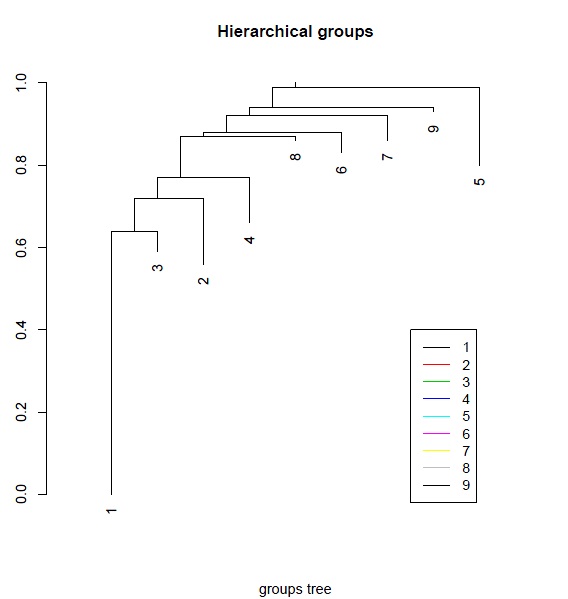}
    }
    \caption{Plots of NPC method results.}
    \label{clusters}
\end{figure}

Figure~\ref{grids} refers to the first form, for three choices of the
geographical area over which the grid of points is constructed.
More precisely, the sets of points for which the computations have been
performed have been obtained as the intersection of  rectangular grids
of sizes $(192 \times 214)$, $(149 \times\,214)$ and $(119\times\,214)$
with the three regions shown on the left side of Figure~\ref{grids},
of different geographical size. This process led then to consider
three non-rectangular grids, comprising  19630, 10696 and 4812  points,
respectively.   The area covered by first grid includes the largest number of
points associated of the seismic events of the data set, while the last one
refers to the area with highest concentration of events.  The right side of
the figure displays the scatter plots of log-slip and log-density of the
points, separately for each cluster. Only clusters labelled No.\,1, 2, 3, 6,
and 8 are considered here, since the other clusters of Figure~\ref{clusters}
have been dropped, for the reason explained earlier.

Even if the selection of the three grids is somewhat subjective, the overall
indication provided from Figure~\ref{grids} provides convincing evidence of
the presence association between the variables under consideration,
specifically clusters labeled with No.\,1, 2, and 6.  This association is
becoming more and more marked as we move down from the first to the last grid,
that is when we focus on the area with greater intensity of events. The type
of association is definitely non-linear, and so admittedly it does not lend
itself to simple interpretation, but it is clearly present, especially so in
the bottom portion of the figure.

\begin{figure}[htb]
    \centering
    \includegraphics[width=10cm,height=8cm]{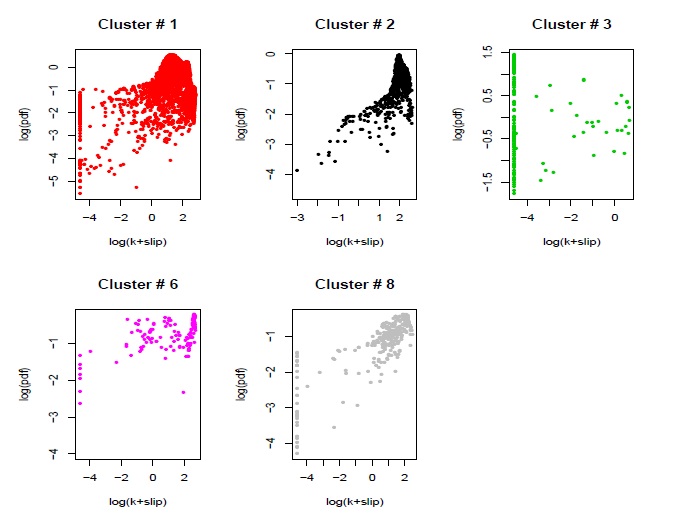}
    \caption{Relationship between log-slip and post-seismicity log-frequency
      for the clusters obtained for the NPC method of the first plot of
      Figure~\ref{clusters} that consider the slip model.
      The  points with $\mathrm{slip}=0$ produce verticals strips of points
      at abscissa $\log(0.01)\approx -4.6$}\label{pdf_slip2}
\end{figure}

Figure~\ref{pdf_slip2} refers instead to the second form of comparison, where
evaluation of log-slip and log-density is performed at the observed location
of events instead of a regular grid of points. Also this figure exhibits some
noteworthy features.  One is that in the red, black, violet, and grey clusters there
exists a clear positive association between log-pdf and log-slip. Hence, the
maximum slip is associated with high frequency of events.  The green cluster
does not display any association,  presumably so because several
events matching with null slip zone where probably have not been involved with
the main earthquake; however the slip produced in this zone is
lower. Pichilemu city ($34\mbox{.}38^{\circ}$S, $72\mbox{.}02^{\circ}$W)
is located in the middle of the red cluster, approximately, which where the
maximum slip is 16.6 meters. The sky-blue cluster correspond to seismic
activity produced by Puyehue volcano eruption (June, 2011). Mocha Island ($38\mbox{.}39^{\circ}$S,
$73\mbox{.}87^{\circ}$W) is located in the bottom of the gray cluster,
which where the maximum slip is 11.9 meters (see Figure~\ref{map}). In the black and gray clusters,
we can see a positive association: the values of log-slip increments in the
measure that the log-pdf increase.

\subsection{Nonparametric analysis of variance}

A positive diagnostic of NPC allows us to implement the Kruskal-Wallis one-way
analysis of variance method (KW; Kruskal \& Wallis, 1952) by ranks that show
the significance existence of slip between the clusters and consider a
confidence level to testing what is the high slip of a cluster over the
others. The KW is a classical method for testing whether samples originate
from the same distribution where the null hypothesis is that the groups from
which the samples originate, have the same median. Since it is a
non-parametric method, the KW test does not assume a normal population or
another distribution, but its purpose is analogous to the one of the classical
analysis of variance for normal populations.  The KW method consider a statistical
test corrected by ties to compute the p-value and, large values of this
test statistic produce the reject of the null hypothesis that the median of
groups are equal. We made use of this method to test whether the clusters
produced by the NPC method are associated to different slip measurements.  The
numerical outcome of the R function {\tt kruskal.test} was $1861.5$ with an
associated $p$-value (below $10^{-15}$), which provides an extreme indication
of heterogeneity of slip among the clusters.

\begin{table}[htb]
\begin{center}
\caption{$P$-values for Wilcoxon test for clusters slip.
}\label{kw}
\vspace{0.5cm}
\begin{tabular}{cccccc}
  \hline
   & red & black & green & violet & gray  \\
  \hline
    red & - & 0  &  0 & 0.0181 & 0  \\
    black & 0  & - &   0 & 0.3497 & 0  \\
    green & 0  & 0  & - & 0   &  0 \\
    violet & 0.0181 & 0.3497 &   0  & -  &  0 \\
    gray & 0  & 0  &   0  & 0  &  -  \\
  \hline
\end{tabular}
\end{center}
\end{table}


To examine where the differences among the groups are, we make use of the
Wilcoxon test (Wilcoxon, 1945).
This test perform individual test between two groups assuming for a null
hypothesis that not exist differences between the two medians. The results are
shown in Table~\ref{kw}.
If each $p$-value is consider isolatedly, there is
only on non-significant comparison at 5\% level, but we must make an allowance
for repeated testing; in this case, $10$ testing procedures have been
performed.  The more classical form of allowance for repeated testing is via
the Bonferroni correction, which here leads to consider the $0.05/10=0.005$
significance level.  Therefore, also the value $0.0181$ must be regarded as
non-significant.

\begin{table}[htb]
\begin{center}
\caption{Summary statistics of the slip variable by cluster and geodetic distance of clusters from the trench.}
\label{stat}
\vspace{0.5cm}
\begin{tabular}{ccccccccc}
\hline
& \multicolumn{5}{c}{Statistics} & \multicolumn{3}{c}{Geodetic Distance}\\
Cluster &  Mean & Min & Max  &  S.D. & N & Min & Max & Mean \\
\hline
red  & 6.200 & 0 & 16.566 &  4.329 & 4165 &  18.34 &326.78 &100.67\\
black    & 7.579 & 0.04 & 13.728 &  2.623 & 950 &  1.89 &173.39  &74.53\\
green  & 0.084 & 0 & 1.945   &  0.331 & 265 & 98.82 &219.64 &147.55\\
violet & 7.572 & 0 &15.043 &  5.752 & 149 &   1.31 &168.93 & 27.56\\
gray   & 4.043 & 0 &11.875  &3.028  &308 &   3.42 &212.91 & 68.75\\
\hline
\end{tabular}
\end{center}
\end{table}

We can see in Table~\ref{stat} that the red cluster representing the
Pichilemu zone has the higher maximum of slip in relation with red
(Constituci\'on zone) and violet (Pichilemu's offshore coast zone) clusters.
With respect the gray (Arauco zone) and green (Rancagua city zone) clusters,
present the lower values of slip and practically, the green cluster does not
present slip.

\begin{figure}[htb]
    \centering
    \includegraphics[width=8cm,height=8cm]{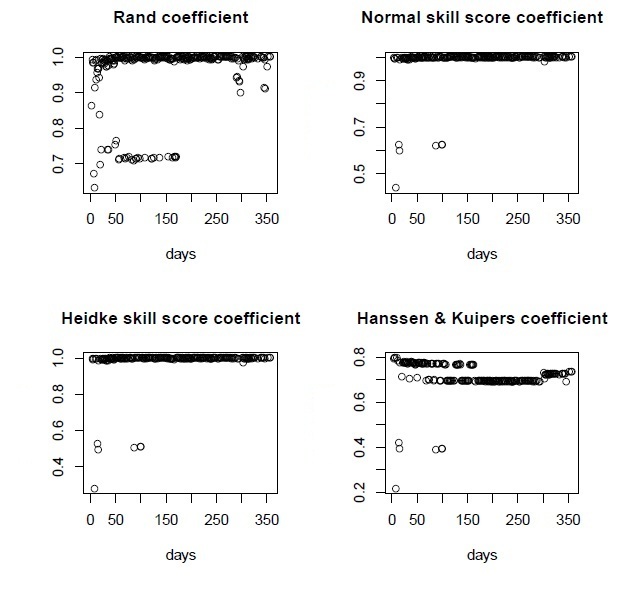}
    \caption{Plots of indexes of NPC results comparisons of data set
        at $t$ day versus $t-1$ day.}\label{coef}
\end{figure}

The clustering estimation can be modified depending of the number of days
after the mega earthquake and the observations involved in each day, so it
may produce unequal results.  Hence, for $t>1$, we compare the results at day $t$
respect to day $t-1$ to compare two alternative partitions of the same set.
In each comparison it is necessary to keep the same number of clusters at day
$t$ and $t+1$. Figure~\ref{coef} shows the consistency of NP method along 200
days since the moment of the great earthquake with values higher
than 0.89 for $HA$ case, 0.95 for $NSS$ and $HSS$ cases, and 0.7
for $HK$ case. In the first 200 days, some compressions between
the clusters estimation at day $t$ versus day $t-1$ produce lower
values of the indexes by the incorporation of one or more new
groups related to the added observations.

\section{Discussion}

The pronounced crustal aftershock activity with mainly normal faulting mechanisms is found in
the Pichilemu region (Far\'ias et al., 2011; Lange et al., 2012, Rietbrock et al., 2012).
Lange et al., (2012) consider the processed events between 15 March
and 30 September 2010 to estimate local magnitudes ($M_l$) in the Pichilemu region, where
those magnitudes are comparable with the SSN magnitudes for large events. Specifically, a crustal
aftershock activity is found in the region of Pichilemu ($\sim34\mbox{.}5^{\circ}$S) where
the crustal events occur in a $\sim30$ km wide region with sharp inclined boundaries
and oriented oblique to the trench. 
On the another hand, the aftershock seismicity parallel to the trench is apparent at 50-120 km distance
perpendicular to the trench (see Table~\ref{stat}). Near $\sim35^{\circ}$S and in the southern part
of the rupture at $\sim38^{\circ}$S occurs significant aftershock activity after the megathrust earthquake.
This seismicity occurs in regions of high coseismic slip (see Table~\ref{stat}). Aftershocks and
coseismic slip of the Maule 2012 earthquake terminate $\sim50$ km south of the prolongation
of the subducting Mocha Fracture zone around $\sim$($73\mbox{.}5^{\circ}$W, $38\mbox{.}5^{\circ}$S),
near of the bottom of gray cluster (see Figure~\ref{map}).

We have proposed an alternative way to clustering the aftershocks seismicity
of the 2010 Maule earthquake $M_W$ 8.8. The nonparametric clustering has shown
to be consistent in the measure that the dairy aftershocks events are added in
the analysis and we present the diagnostic tools to illustrate this feature.
We using a nonparametric kernel method to fit the high empirical aftershock
frequency, which were highly correlated with the used coseismic slip model. 
Our findings can be explored further by considering an extended data
set, including the events with delayed effect, and modeling the
relationship of high coseismic slip areas and aftershock clusters.
Also, this catalogue should be considered to the study of the behavior of a typical aftershock
sequence, to identify outliers and to classify sequences into groups
exhibiting similar aftershock behavior (Schoenberg \& Tranbarger, 2008; Schoenberg et al., 2006).
Finally, this analysis should be considered in the attempt to help
identification of an increasing risk of occurrence of another great earthquake.

\section*{Acknowledgements}

We like to thank Giovanna Menardi for useful comments and suggestions in connection with the pdfCluster methodology;
to Eduardo Contreras-Reyes and Marcos Moreno for useful comments, suggestions and computational help; and Hector
Massone for seismicity data base used in this article.

\section*{References}

\newenvironment{reflist}{\begin{list}{}{\itemsep 0mm \parsep 1mm
\listparindent -7mm \leftmargin 7mm} \item \ }{\end{list}}
\baselineskip 16pt
{\small

\refmark Azzalini, A., Torelli, N., 2007. Clustering via nonparametric
density estimation. \emph{Stat. Comput.} 17, 71-80.

\refmark Azzalini, A., Menardi, G., Rosolin, T., 2011. R package
\emph{pdfCluster}: Cluster analysis via nonparametric density estimation
(version 0.1-13). URL {\tt http://cran.r-project.org/package=pdfCluster}

\refmark Botev, Z.I., Grotowski, J.F., Kroese, D.P. (2010). Kernel Density
Estimation Via Diffusion. {\it Ann. Stat.} 38, 5, 2916-2957.

\refmark Bowman, A.W., Azzalini, A., 1997. \emph{Applied Smoothing
  Techniques for Data Analysis: The Kernel Approach with S-Plus Illustrations}.
Oxford University Press, Oxford.

\refmark Brier, G.W., Allen, R.A., 1952. Verification of weather forecast.
\emph{Compendium of Meteorology}, Boston, \emph{Amer. Meteor. Soc.}, 841-848.

\refmark Contreras-Reyes, E., Carrizo, A., 2011. Control of high oceanic
features and subduction channel on earthquake ruptures along the Chile-Per\'u
subduction zone. \emph{Phys. Earth Planet. Inter.} 186, 49-58.

\refmark Contreras-Reyes, J.E., Arellano-Valle, R.B., 2012. Kullback-Leibler
divergence measure for Multivariate Skew-Normal Distributions. 
\emph{Entropy} 14, 9, 1606-1626. doi:10.3390/e14091606.

\refmark Darwin, C., 1851. Geological Observations on Coral Reefs, Volcanic Islands and on
South America. \emph{Londres, Smith, Elder and Co}, pp. 768.

\refmark Dasgupta, A., Raftery, A.E., 1998. Detecting features in spatial
point processes with cluster via model-based clustering.
\emph{J. Amer. Statist. Assoc.} 93, 294-302.

\refmark Delouis, B., Nocquet, J.M., Vallee, M., 2010. Slip distribution of
the February 27, 2010 Mw = 8.8 Maule Earthquake, central Chile,
from static and high-rate GPS, InSAR, and broadband teleseismic data.
\emph{Geophys. Res. Lett.} 37, L17305.

\refmark Far\'ias, M., Comte, D., Roecker, S., Carrizo, D., Pardo, M., 2011. Crustal
extensional faulting triggered by the 2010 Chilean earthquake: The Pichilemu Seismic
Sequence. \emph{Tectonics} 30, TC6010.

\refmark Hansen, A.W., Kuipers, W.H.A., 1965. On the relationship between the frequency of rain and various
meteorological parameters. Koninklijke Nederlands Meteorologisch Institut, \emph{Meded. Verhand.}  81, 2-15.

\refmark Hubert, L., Arabie, P., 1985. Comparing Partitions. \emph{J. Classif.} 2, 193-218.

\refmark Kaufman, L., Rousseeuw, P.J., 1990. \emph{Finding Groups in
  Data: An Introduction to Cluster Analysis}. New York: Wiley.

\refmark Kruskal, W., Wallis, W.A., 1952. Use of ranks in one-criterion
variance analysis. \emph{J. Amer. Statist. Assoc.} 47, 260, 583-621.

\refmark Lange, D., Tilmann, F., Barrientos, S., Contreras-Reyes, E., Methe, P., Moreno, M., Heit, B.,
Bernard, P., Vilotte, J., Beck, S., 2012. Aftershock seismicity of the 27 February 2010 Mw 8.8
Maule earthquake rupture zone. \emph{Earth Planet. Sci. Lett.} 317-318, 413-425.

\refmark Lay, T., Ammon, C.J., Kanamori, H., Koper, K.D., Sufri, O., Hutko, A.R., 2010.
Teleseismic inversion for rupture process of the 27 February 2010 Chile (Mw= 8.8)
earthquake. \emph{Geophys. Res. Lett.} 37, L13301.

\refmark Lorito, S., Romano, F., Atzori, S., Tong, X., Avallone, A., McCloskey, J., Cocco, M., Boschi, E.,
Piatanesi, A., 2011. Limited overlap between the seismic gap and coseismic slip of
the great 2010 Chile earthquake. \emph{Nat. Geosci.}  4, 173-177.

\refmark Menardi, G., 2010. Density-based silhouette diagnostics for
clustering methods. \emph{Stat. Comput.} 21, 295-308.

\refmark Moreno, M., Rosenau, M., Oncken, O., 2010. Maule earthquake slip
correlates with pre-seismic locking of Andean subduction zone. \emph{Nature}
467, 198-202.

\refmark Moreno, M., Melnick, D., Rosenau, M., Baez, J., Klotz, J., Oncken, O., Tassara, A., Chen, J., Bataille, K.,
Bevis, M., Socquet, A., Bolte, J., Vigny, C., Brooks, B., Ryder, I., Grund, V., Smalley, B., Carrizo, D.,
Bartsch, M., Hase, H., 2012. Toward understanding tectonic control on the
Mw 8.8 2010 Maule Chile earthquake. \emph{Earth Planet. Sci. Lett.},
321-322, 152-165.

\refmark Pollitz, F.F., Brooks, B., Tong, X., Bevis, M.G., Foster, J.H., Bürgmann, R., Smalley, R.J.,
Vigny, C., Socquet, A., Ruegg, J.-C., Campos, J., Barrientos, S., Parra, H., Baez Soto, J.-C.,
Cimbaro, S., Blanco, M., 2011. Coseismic slip distribution of the February 27, 2010 Mw 8.8 Maule,
Chile earthquake. \emph{Geophys. Res. Lett.} 38, L09309.

\refmark R Development Core Team, 2012. \emph{R: A Language and Environment
  for Statistical Computing}. R Foundation for Statistical Computing, Vienna,
Austria.  ISBN 3-900051-07-0, URL http://www.R-project.org.

\refmark Rietbrock, A., et al., 2012.
Aftershock seismicity of the 2010 Maule Mw=8.8, Chile, earthquake: Correlation between co-seismic slip models
and aftershock distribution? \emph{Geophys. Res. Lett.} 39, L08310.

\refmark Rousseeuw, P.J., 1987. Silhouettes: A graphical aid to the interpretation and validation of
cluster analysis. \emph{J. Comput. Appl. Math.} 20, 53-65.

\refmark Ruegg, J., Rudloff, A., Vigny, C., Madariaga, R., de Chabalier, J.B., Campos, J., Kausel, E.,
Barrientos, S., Dimitrov, D., 2009. Interseismic strain accumulation measured
by GPS in the seismic gap between Constituci\'on and Concepci\'on in Chile.
\emph{Phys. Earth Planet. Inter.} 175, 78-85.

\refmark Schoenberg, F.P., Brillinger, D.R., Guttorp, P., 2006.
\emph{Point Processes, Spatial-Temporal}. Encyclopedia of Environmetrics.

\refmark Schoenberg, F.P., Tranbarger, K.E., 2008. Description of earthquake aftershock sequences using
prototype point patterns. \emph{Environmetrics} 19, 271-286.

\refmark Vigny, C., Socquet, A., Peyrat, S., Ruegg, J.-C., Métois, M., Madariaga, R., Morvan, S.,
Lancieri, M., Lacassin, R., Campos, J., Carrizo, D., Bejar-Pizarro, M., Barrientos,
S., Armijo, R., Aranda, C., Valderas-Bermejo, M.-C., Ortega, I., Bondoux, F., Baize,
S., Lyon-Caen, H., Pavez, A., Vilotte, J.P., Bevis, M., Brooks, B., Smalley, R., Parra,
H., Baez, J.-C., Blanco, M., Cimbaro, S., Kendrick, E., 2011. The 2010 Mw 8.8 Maule Mega-Thrust
Earthquake of Central Chile, Monitored by GPS. \emph{Science} 332
(6036), 1417-1421.

\refmark Wilcoxon, F., 1945. Individual comparisons by ranking methods.
\emph{Biometrics} 1, 80-83.

\refmark Zhu, Y., 2013. Nonparametric density estimation based on the
truncated mean. \emph{Stat. Probabil. Lett.} 83, 445-451.

}
\renewcommand{\baselinestretch}{1}

\end{document}